# Sub-nanosecond all-optically reconfigurable photonics in optical fibres

**Kunhao Ji**[1 *], **David J. Richardson**[1,2], **Stefan Wabnitz**[3], **Massimiliano Guasoni**[1 *]

1. Optoelectronics Research Centre, University of Southampton, Southampton SO17 1BJ, United Kingdom
2. Microsoft (Lumenisity Limited), Unit 7, The Quadrangle, Abbey Park Industrial Estate, Romsey, SO51 9DL, United Kingdom
3. Department of Information Engineering, Electronics and Telecommunications (DIET), Sapienza University of Rome, 00184 Rome, Italy

**We introduce a novel all-optical platform in multimode and multicore fibres. By using a low-power probe beam and a high-power counter-propagating control beam, we achieve advanced and dynamic control over light propagation within the fibres. This setup enables all-optical reconfiguration of the probe, which is achieved by solely tuning the control beam power. Key operations such as fully tuneable power splitting and mode conversion, core-to-core switching and combination, along with remote probe characterization, are demonstrated at the sub-nanosecond time scale. Our experimental results are supported by a theoretical model that extends to fibres with an arbitrary number of modes and cores. The implementation of these operations in a single platform underlines its versatility, a critical feature of next-generation photonic systems. These results represent a significant shift from existing methods that rely on electro-optical or thermo-optical modulation for tunability. They pave the way towards a fast and energy-efficient alternative through all-optical modulation, a keystone for the advancement of future reconfigurable optical networks and optical computing. Scaling these techniques to highly nonlinear materials could underpin ultrafast all-optically programmable integrated photonics.**

## INTRODUCTION

The ability to manipulate light by light within optical fibres represents a pivotal advance, both for the development of new photonic technologies and the exploration of novel physical phenomena. Groundbreaking all-optical devices and applications have been developed in single-mode fibres, including optical amplifiers, signal regeneration, polarization control, sensing and logical operations.[1–7]

The recent renewed interest in multimode fibres (MMFs) and multicore fibres (MCFs), driven by the need for high-speed communication systems based on space-division multiplexing (SDM) [8,9], has sparked attention to complex nonlinear multimode processes that have no counterpart in the single-mode platform, and whose comprehension is still in the early stages [10–13], paving the way for new methods of all-optical control of light.

In the framework of all-optical control, we can differentiate between self-organization and external control. Self-organization occurs when an intense light beam reshapes its own dynamics, owing to the substantial nonlinearity induced by its large peak power. Beam self-cleaning[14,15], self-switching[16,17], self-coherent combination[18] and self-repolarization processes [19,20] induced by Kerr nonlinearity fall into this category. Conversely, external control occurs when the dynamics of a probe beam are controlled by an external independent control beam through their mutual nonlinear interaction.

When both the probe and control beam are relatively intense, their nonlinear interaction may exhibit robust modal attraction [21–24] or even rejection dynamics [25], as recently demonstrated in multimode systems. In contrast, when the probe beam operates in a low-power (linear) regime, substantially different dynamics emerge, where the control beam induces a periodic optical grating inscribed in the fibre. Optically induced gratings, so far limited to bimodal systems, have been exploited to implement partial mode conversion of the probe beam [26–28].

In this work we propose a counter-propagating probe-control beam scheme in MMFs and MCFs with arbitrary number *N* of modes or cores. This setup allows the simultaneous phase-matching of several interaction processes between a low-power, forward probe and an intense backward control beam (BCB), regardless of the fibre parameters, thus harnessing the full potential of multimode dynamics. By leveraging a robust setup for accurate mode coupling and mode decomposition, we provide an experimental demonstration of several new compelling all-optical operations in MMFs and MCFs, which include fast and fully tuneable mode conversion and power splitting, selective core-to-core switching and combining, as well as the remote characterization of the probe beam, as illustrated in Fig. 1.

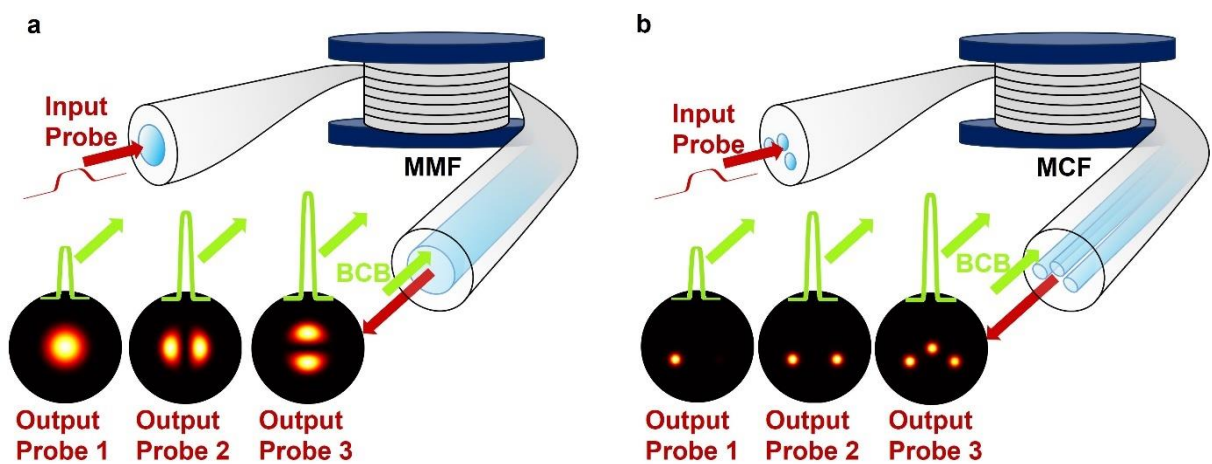

**Fig. 1: Illustration of all-optically reconfigurable photonics in optical fibres. a.** A low-power probe beam (red colour) and a high-power counter-propagating backward control beam (BCB, green colour) are injected at the two opposite ends of a multimode fibre. The BCB is coupled over a suitable combination of modes. A specific output probe on demand can be obtained by solely adjusting the BCB power. In this example, 3 different BCB intensities lead to an output probe coupled over 3 distinct fibre modes (see Output Probe 1,2,3). **b.** Same as panel a, but in the case of a multicore fibre with 3 cores. In this example, by tuning the BCB intensity, the output probe is either fully readdressed over a single core (Output Probe 1), or equally split over 2 (Output Probe 2) or 3 cores (Output Probe 3). The ability to manipulate the probe can be exploited to implement power splitters, mode converters and core-to-core switchers with all-optical reconfiguration at the sub-nanosecond scale.

These outcomes reveal a novel all-optical control mechanism for configuring the modal state and optical pathways in MMFs and MCFs, enabling novel functionalities for future smart and adaptive optical systems. This lays the groundwork for a new all-optically reconfigurable photonics in optical fibres and beyond[29].

**RESULTS**

**Probe-control beam interaction.** A forward probe signal and a BCB are counter-propagating in a fibre supporting *N* spatial modes. Their spatio-temporal evolution is described by a system of coupled nonlinear Schrödinger equations[25](CNLSEs, see Methods). The counter-propagating setup offers some key advantages with respect to a standard co-propagating configuration. By physically separating probe and BCB, and injecting them on the opposite ends of the fibre, the probe-BCB intermodal interactions in the CNLSEs can be all simultaneously phase-matched (see Methods), which allows for an energy exchange among all the probe modes, rather than a single pair of phase-matched modes. In addition, the separation of probe and BCB allows for the implementation of remote sensing operations, hence, to investigate the properties of the fibre and/or of the probe, even when the latter is inaccessible.

In a recent work [25] we analysed the case where both probe and BCB operate in a strongly nonlinear regime, which exhibits robust mode attraction or rejection states, irrespectively of the initial state of the probe. In this study, we explore a different scenario, where the probe power is weak (linear regime), whereas the BCB is in a strongly nonlinear propagation regime. This leads to peculiar new dynamics, fundamentally different from mode attraction and rejection. After some algebra and appropriate transformations, we recast the CNLSEs into the following system of equations (see Methods):

$$\mathcal{F}_{out} = M \mathcal{F}_{in} \qquad (1)$$

where $\mathcal{F}_{in}$ and $\mathcal{F}_{out}$ are vectors of length $N$ whose entries $f_{in,n}$ and $f_{out,n}$ indicate the amplitude of the electric field of the probe mode $n$ at the input and output of the fibre, respectively, whereas $M$ is a $N \times N$ matrix whose elements are defined by the BCB mode state, along with the nonlinear Kerr coefficients of the fibre and the modal propagation constants (see Methods).

Besides describing the mode dynamics in MMFs and MCFs, equation (1) also characterizes the core-to-core interaction in MCFs, following the identification of the transformation matrix $T$ that links the electric field of the MCF modes to the electric fields in the individual cores, namely:

$$\mathcal{F}_{\substack{c-in \\ c-out}} = T \mathcal{F}_{\substack{in \\ out}} \qquad (2)$$

where $\mathcal{F}_{c-in}$ and $\mathcal{F}_{c-out}$ are vectors of length $N$ whose entries $f_{c-in,n}$ and $f_{c-out,n}$ indicate the amplitude of the electric field of the probe in core $n$ at the input and output of the fibre, respectively.

Vector $\mathcal{F}_{in}$ ($\mathcal{F}_{out}$) describes the input (output) probe mode state, which includes information on both the input (output) mode power distribution $|\mathcal{F}_{in}|^2$ ($|\mathcal{F}_{out}|^2$) and the relative phase between the modes at the input (output) of the fibre. Likewise, vectors $\mathcal{F}_{c-in}$ and $|\mathcal{F}_{c-in}|^2$ ($\mathcal{F}_{c-out}$ and $|\mathcal{F}_{c-out}|^2$) represent the input (output) probe core state and power distribution, respectively. Additionally, we can define the BCB mode state and power distribution in a similar manner. The probe and BCB mode states and power distributions exhibit an important relationship, as discussed in the Methods section.

The importance of equations (1) and (2) lies in their establishment of a direct relationship between the input and output probe states. By appropriately configuring the BCB, we shape matrix $M$ to achieve a mode or core state on demand in the output probe. In other words, we implement all-optical reconfiguration of the output probe.

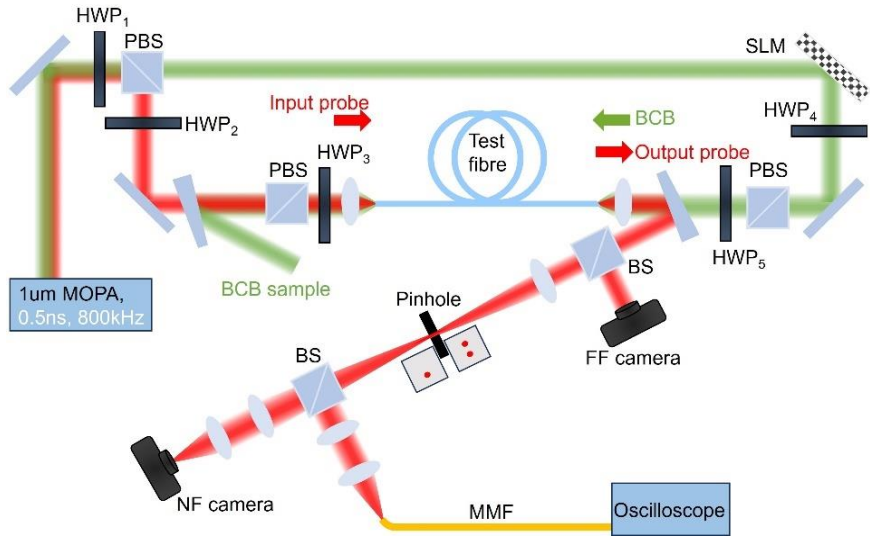

**Fig. 2: Schematic of the experimental setup.** The input probe and BCB are split from a master oscillator power amplifier (MOPA) and coupled to the opposite ends of the test fibre. The MOPA delivers 0.5 ns pulses with peak power up to 30 kW (12 W average power at 800 kHz repetition rate), therefore enabling a significant level of nonlinearity in the fibres under test. Polarization beam splitters (PBS) and half-wave-plates (HWP$_{1-5}$) are used to tune independently the input probe and BCB power and polarization. A near-field (NF) and a far-field (FF) camera measure the near and far field images used in our mode decomposition algorithm. The field at the output of each core of MCFs can be isolated via a pinhole and its temporal dynamic is monitored at the oscilloscope. SLM=spatial light modulator; BS=beam splitter.

**Application 1: Tuneable mode manipulation.** A first key application of our platform is the tuneable mode manipulation of the probe. In our experiments (see Fig. 2 and Methods), we used a bimodal fibre and two three-mode fibres (see Supplementary Information 1). For simplicity of illustration, we present here the results obtained in the bimodal fibre. A summary of the outcomes with three-mode fibres is reported in the Supplementary Information 2, including a highly nonlinear fibre that relaxes substantially the power requirements on the BCB.

The results in Fig.3 demonstrate a tuneable all-optical mode conversion, where any arbitrary power ratio between the two guided modes can be achieved by solely adjusting the BCB power. Three distinct instances are shown that highlight the extent of the precision in manipulating the probe mode distribution. Indeed, for a given input mode state of the probe, we can configure the BCB in order to achieve either full mode conversion of the output probe (Fig. 3d), partial mode conversion (Fig. 3e), or conversion annihilation, thus making the output probe mode distribution insensitive to the probe-BCB interaction (Fig. 3f).

Remarkably, our experimental results in Fig. 3d-f closely align with the theoretical predictions derived from equation (1), validating the precision of our mode decomposition setup. Note that the relative polarization between probe and BCB may serve as an additional parameter for controlling the probe dynamics (see Supplementary Information 3).

As previously anticipated, and further elaborated in the Supplementary Information 4, the underlying spatio-temporal dynamics present some fundamental differences when compared to the case where both beams counter-propagate in a high-power regime.[25] Firstly, the BCB mode distribution is unaffected by the nonlinear dynamics: therefore, it remains unchanged during propagation. Moreover, as is evident from Fig. 3, the output probe evolution versus BCB power follows a periodic (sinusoidal-like) trend, rather than converging to or rejecting a specific mode state.

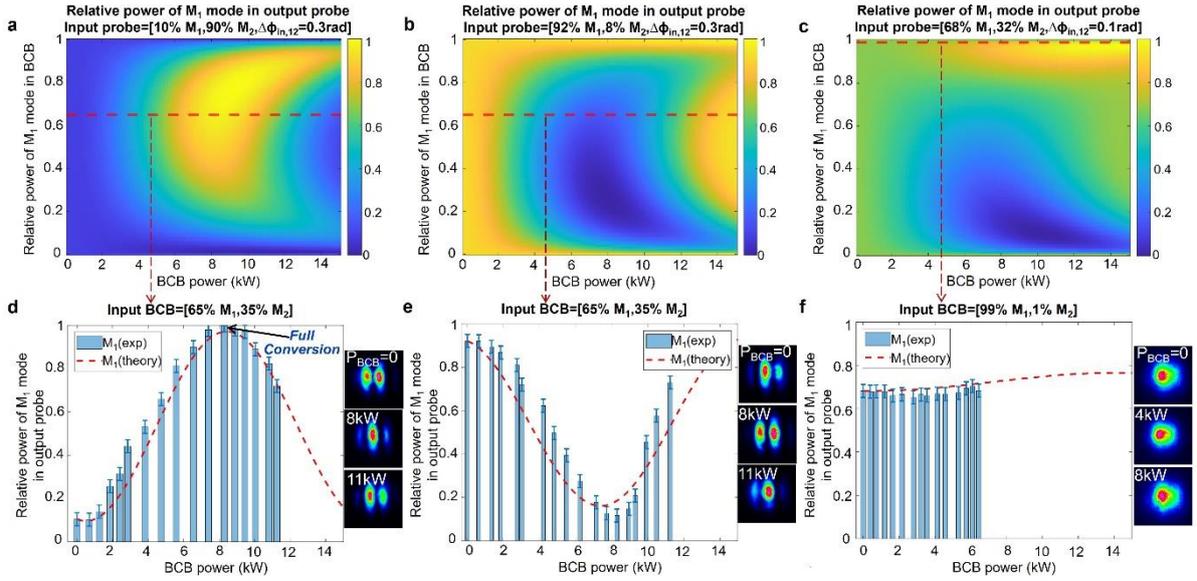

**Fig. 3: Tuneable mode manipulation.** Results in the bimodal fibre. This fibre is 0.4 meter long and supports one even mode $M_1$ and one odd mode $M_2$ (see Supplementary Information 1). **a-c.** Theoretical 2D maps of the output probe mode distribution computed from equation (1). The maps show the output probe power fraction coupled to mode $M_1$ versus the BCB total peak power (horizontal axis) and BCB mode distribution (vertical axis, indicating the fraction of BCB power coupled to mode $M_1$). These maps indicate how to set the BCB in order to manipulate the output probe, ensuring it reaches the desired mode distribution. The maps correspond to 3 examples with different input probe mode states, which are reported at the top of each panel. For example, in panel a the input probe mode state is characterized by 10% power on mode $M_1$, 90% on mode $M_2$, and a relative phase $\Delta\phi_{in,12}$ between the two modes of 0.3 rad. **d-f.** Experimental (exp) and theoretical (theory) results for the same input probe mode states as panels a-c, but with a fixed BCB mode distribution (indicated at the top of each panel and corresponding to the red-dashed lines in panels a-c). Arbitrary output probe mode distribution can be achieved by tuning the BCB power. Specifically, in panel d, full conversion to mode $M_1$ is achieved when the BCB peak power is ~ 8 kW. In contrast, the BCB in f is configured such that it results in almost no variation of the output probe mode distribution. The insets in panels d-f show the far-field intensities of the output probe for different values of BCB peak power $P_{BCB}$. Error bars of ±3% are added to the measured relative power of $M_1$ mode, which represents the estimated uncertainty of our mode decomposition algorithm.

**Application 2: Tuneable power splitting, core-to-core switching and combining.** A significant feature of our setup lies in the possibility to manipulate the core-to-core energy exchange in MCFs. Note that while the mode distribution remains largely unaffected by linear coupling in the short fibres under test, core-to-core linear coupling takes place over a much shorter length scale instead. More generally, a complex interplay occurs between linear core-to-core coupling and nonlinear coupling between probe and BCB. A relatively simple, yet instructive case study is a dual-core fibre (DCF) of length $L$ where the input probe with total power $P_p$ is coupled to a single core, say core *1*, and the BCB with total power $P_{BCB}$ is coupled to a single mode, say mode *1*. In this instance the output probe power $\left|f_{c-out,n}\right|^2$ in the 2 cores (n={1,2}), computed from equations (1) and (2), reads as:

$$\left|f_{c-out,1}\right|^2 = P_p cos^2(\pi L/L_b + \Delta\gamma\, P_{BCB}\, L)$$

$$\left|f_{c-out,2}\right|^2 = P_p sin^2(\pi L/L_b + \Delta\gamma\, P_{BCB}\, L) \qquad (3)$$

where $L_b = 2\pi/|\beta_1 - \beta_2|$ is the beat-length between the two modes of the DCF having propagation constants $\beta_{1,2}$, and $\Delta\gamma = \gamma_{11} - \gamma_{12}$ is the difference between the Kerr coefficients $\gamma_{11}$ and $\gamma_{12}$. When BCB is off ($P_{BCB}$=0), the probe undergoes core-to-core energy exchange over a distance as short as $L_b$ (typically a few millimetres). Modal beat-lengths are severely affected by fibre perturbations, such as local bending and temperature fluctuations, and are therefore difficult to estimate. However, equation (3) highlights a crucial point. Irrespectively of the beat-length $L_b$, which may even be unknown, the output probe power in the two cores is fully tuneable by adjusting the BCB power $P_{BCB}$, enabling any arbitrary splitting ratio. Importantly, this finding is generalizable to different fibre parameters and input conditions.

Our experimental results, shown in Fig. 4a, confirm this scenario. The input probe launch condition was adjusted such that, with the BCB off, the output probe power was fully coupled to a single core. By introducing the BCB and tuning its peak power between 0 and 9 kW, we achieved any arbitrary power ratio X/(100–X) between the two output cores.

Additional key applications can be envisaged and demonstrated with our platform. As shown in Fig. 4b, the power of the output probe, which in this case is relatively uniform in the 2 cores when BCB is off (power ratio core 1/core 2=35/65), can be combined into core 1 when the BCB peak power is set to 11 kW. Furthermore, core-to-core switching is depicted in Fig. 4c, where the output power transitions from one core to another at a BCB power of ~10 kW. Note that, in this case, the switching power ratios (from 15/85 to 85/15) are constrained by the available coupled BCB peak power, which is <12 kW in our experiments in the DCF. Approximately 18 kW of BCB peak power would be required for complete 0/100 to 100/0 switching (indeed 9kW allows 100/0 to 50/50 splitting, see Fig. 4a).

The applications highlighted above can be controlled at an extremely fast rate through the BCB. Figs. 4d,e illustrate the sub-nanosecond modulation of the core-to-core power ratio. The temporal evolution of the output probe power in the 2 cores, measured via an oscilloscope, is displayed. A single 0.5 ns BCB pulse shifts the core-to-core power ratio at the DCF output from 35/65, when the BCB is off, to 65/35 when the BCB peak power is 5 kW. The switching time is determined by the BCB pulse width. Although in our experiments the BCB pulse width is 0.5 ns, the simulations in Supplementary Information 5 indicate that the switching time could be reduced to picosecond levels. These results pave the way for the development of all-optically controlled core-to-core switchers, leading to the pioneering idea of all-optically programmable photonics. In particular, the DCF with BCB control could serve as basic unit (2x2 optical gate) for reconfigurable wide matrices[30,31], enabling fully optical ultrafast operations.

In this framework, exploring complex multicore systems is compelling. A single BCB beam could enable core-to-core switching, splitting or combining with N>2 cores. These systems are more sensitive to weak variations in fibre parameters than the DCF. Our generalized solutions in equations (1) and (2), which effectively describe the modal dynamics, would require precise knowledge of the relative differences among intermodal beat-lengths in order to describe the core-to-core dynamics equally well. However, these differences are susceptible to perturbations, therefore their estimation is challenging. Consequently, in our experiments we manually adjust the BCB mode state to find the optimal configuration that enables the desired control over the probe beam.

Despite these challenges, our theoretical model remains invaluable, suggesting intriguing scenarios. For instance, the simulation results in Fig. 5a indicate that, with sufficient BCB power, coherent combination or equal splitting could be achieved in a three-core-fibre (TCF). Preliminary experimental tests support the feasibility of these outcomes. Although the coupled BCB power is significantly lower than the simulated values, preventing full power rerouting in each core, nevertheless we could split the power evenly across the 3 cores (Fig. 5b), combine power from 2 cores into a single core (Figs. 5c) or swap the power among selected cores (Fig. 5d).

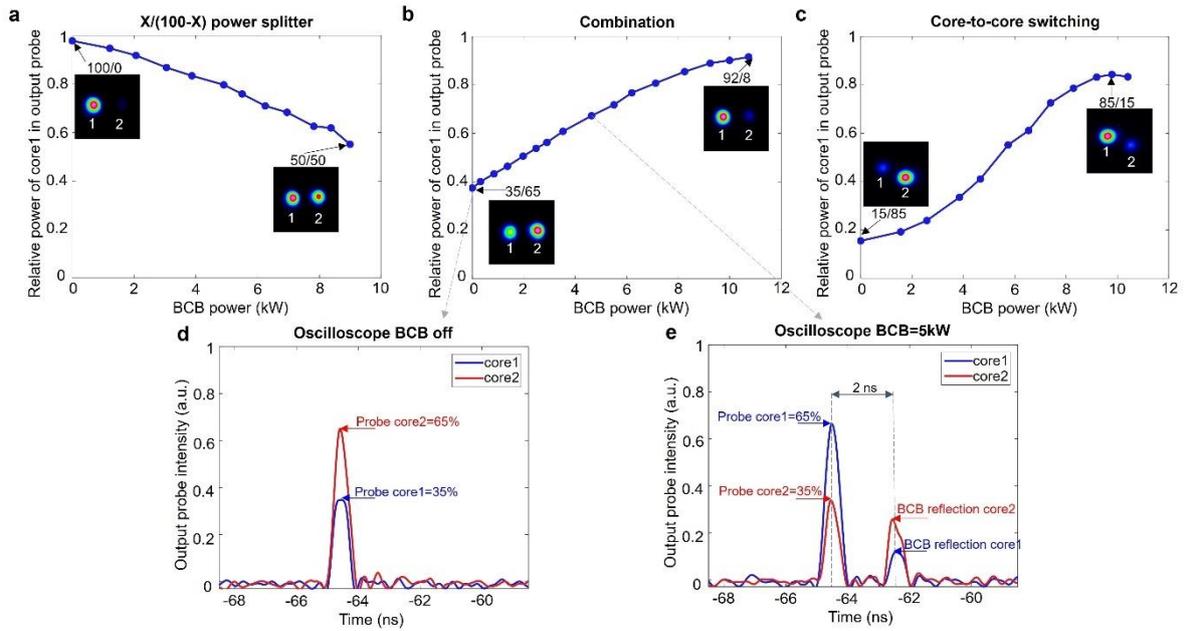

**Fig. 4: Tuneable reconfiguration in dual-core fibre.** Three different instances are shown. The insets show the near-field intensities of the output probe at each core. **a.** The input probe launch condition is optimized such that the output probe power is entirely in core 1 when the BCB is off (power ratio core1/core2 = 100/0). After having appropriately fixed the BCB mode state, we increase the BCB peak power from 0 to 9 kW. We then observe that the core-to-core power ratio of the output probe transitions gradually from 100/0 to 50/50, enabling an all-optical, fully tuneable X/(100-X) power splitting. **b.** Differently from panel a, in this case the output probe core distribution is relatively uniform when the BCB is off (power ratio core1/core2 = 35/65). The output probe is then progressively redirected into core 1 as the BCB power increases, achieving an all-optically controlled combination. At 11 kW of BCB peak power, 92% of the output probe power is in core 1 (power ratio core1/core2 = 92/8). We estimate that full combination (100/0) could be achieved at ~14 kW peak BCB power (not available). **c.** In this example, the output power ratio goes from 15/85 when BCB is off to 85/15 when the BCB peak power is ~10 kW. Full switching (0/100 to 100/0) could be achieved with ~18 kW BCB peak power (not available). **d.** Temporal evolution of output probe power at the two cores measured by the oscilloscope when the BCB is off (power ratio core1/core2 = 35/65). **e.** Temporal evolution of output probe power at the two cores measured by the oscilloscope at 5 kW BCB peak power. The power ratio shifts to 65/35. The oscilloscope also detects the BCB reflection, with the 2 ns delay corresponding to the time of flight of light in the fibre.

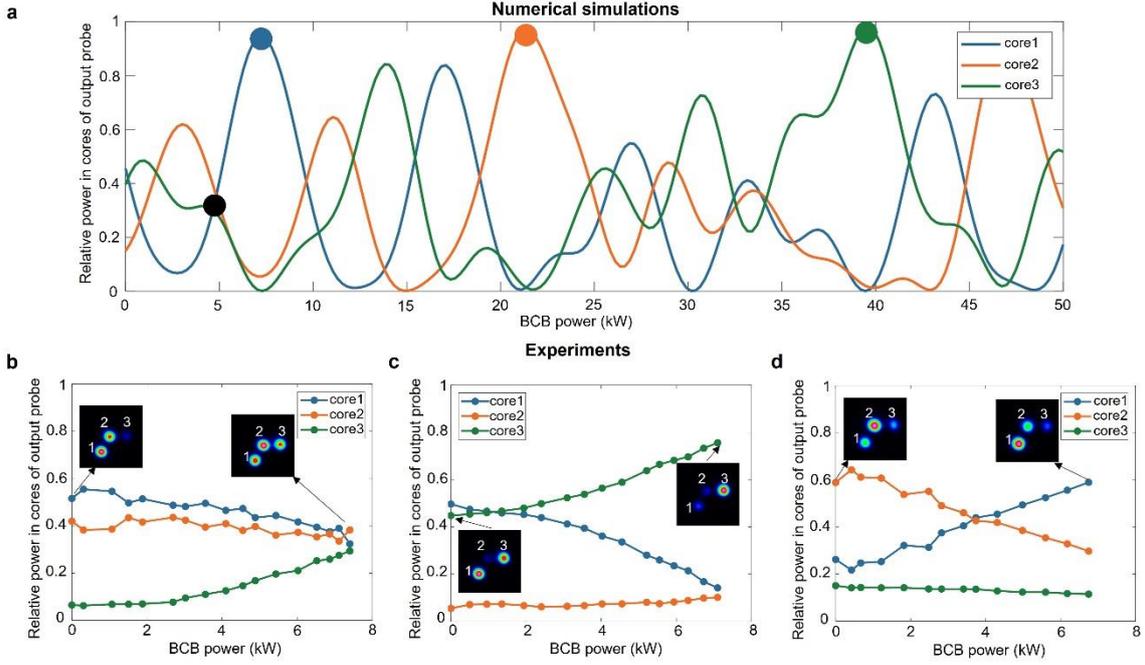

**Fig. 5: Tuneable reconfiguration in three-core fibre.** Our ability to implement all-optical probe reconfiguration extend to fibres with more than 2 cores. This figure illustrates all-optical operations in a 0.4m long TCF. The insets show the near-field intensities of the output probe at each core. **a**. Output probe core distribution simulated via equations (1) and (2), with linear and nonlinear coefficients estimated from the fibre parameters (see Supplementary Information 1). In this simulation, the BCB mode state is as follows: 5% of power in mode 1, 30% in mode 2, 65% in mode 3, and all modes in-phase. The probe power can be arbitrary low. By adjusting the BCB power from 0 to 50 kW we can either equalize the output probe power in the 3 cores (see black spot) or combine most of the output probe power in core 1 (blue spot), core 2 (red spot) or core 3 (green spot). **b-d.** Experimental results in the TCF. Each panel corresponds to different launch conditions of the input probe. In each case, the BCB is optimized to achieve relevant operations for a BCB peak power of ~7kW (maximum power that we can couple to the TCF). In panel b, the output probe is almost equally split across the 3 cores. In panel c, the probe is mainly redirected to a single core (core 3). In panel d, we achieve power swapping between core 1 and core 2.

**Application 3: Probe Remote Characterization.** Our counter-propagating setup could have significant applications in remote sensing, enabling the investigation of fibre or input probe features through the analysis of the output probe's response to the BCB. For instance, consider estimating the input probe mode state into a MMF with $N$ modes and of length $L$. Assuming weak mode coupling, as in few-meter long polarization-maintaining fibres, the probe mode distribution remains constant during propagation when the BCB is off, namely $|f_{in,n}|^2 = |f_{out,n}|^2$. Additionally, there is a direct relationship between the output and input relative phases: $\Delta\phi_{in,1n} = \Delta\phi_{out,1n} - \Delta\beta_{1n}L$, where $\Delta\phi_{in,1n}$ ($\Delta\phi_{out,1n}$) is the input (output) relative phase between mode $n$ and a reference mode, here mode 1, whereas $\Delta\beta_{1n}L$ is the phase delay due to the differential propagation constant $\Delta\beta_{1n}$ between modes 1 and $n$. Consequently, the input probe mode state $\left(|f_{in,n}|^2, \Delta\phi_{in,1n}\ 1 \leq n \leq N\right)$ could, in principle, be inferred by computing the output probe mode state $\left(|f_{out,n}|^2, \Delta\phi_{out,1n}\ 1 \leq n \leq N\right)$ via mode decomposition at the fibre output. However, as previously mentioned, $\Delta\beta_{1n}$ is highly sensitive to fibre perturbations, and even a small error in $\Delta\beta_{1n}$ would result in an unreliable estimate of $\Delta\phi_{in,1n}$.

An efficient solution to this problem consists of analysing the probe's response to the BCB. Indeed, the output probe mode distribution depends on the input relative phases $\Delta\phi_{in,1n}$. Thus, to determine the latter, we computed the theoretical mode distribution for various values of relative phases, and identified the optimal least-squares values that best align with the experimental data. Fig. 6 illustrates three distinct cases where the input phase $\Delta\phi_{in,12}$ is successfully retrieved in a bimodal fibre, even when there is a significant power imbalance between the modes. Notably, in MCFs, once the relative phases $\Delta\phi_{in,1n}$ are recovered, one may estimate through equation (2) the input probe core distribution and relative phase at each core. Moreover, a similar approach could be used to estimate simultaneously both the input probe properties and unknown fibre parameters (e.g. Kerr coefficients, average linear mode coupling) through multivariate estimation analysis.

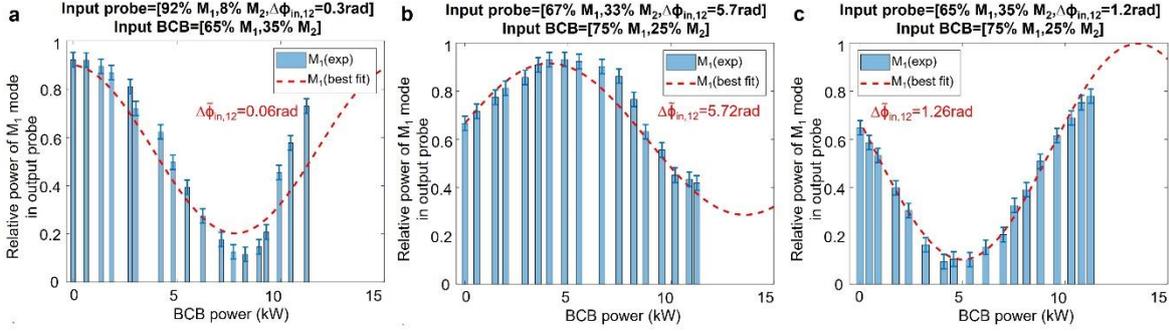

**Fig. 6: Remote characterization of the input probe.** Experimental results (bars) and corresponding best theoretical fits (red-dashed lines) showing the output probe power fraction coupled to mode $M_1$ versus BCB peak power in a 0.4-m long bimodal fibre (DCF, see Supplementary Information 1). Panels a-c correspond to different input probe mode states and BCB mode distributions, measured experimentally and reported on the top of each panel. The best theoretical fit is calculated from equation (1), assuming the same input probe and BCB relative powers and optimizing the input probe relative phase to minimize the least squares difference with experimental data. Note that in all the 3 cases the estimated optimal least-squares value $\Delta\widetilde{\phi}_{in,12}$ (0.06 rad, 5.72 rad, 1.26 rad in panels a, b, c respectively) is close to the measured $\Delta\phi_{in,12}$ (0.3 rad, 5.7 rad, 1.2 rad in panels a, b, c respectively). This demonstrates our ability to detect from remote the relative phase of the input probe modes by analysing the output probe response to the BCB. Note that the larger error in panel a is due to the large power imbalance among the two input probe modes (92% and 8%, respectively).

## DISCUSSION

Our work presents a platform based on a counterpropagating probe-control beam setup in multimode and multicore fibres, which incorporates precise mode decomposition. This setup allows all-optical manipulation of the probe and enables novel key operations at the sub-nanosecond time scale, including fully tuneable mode conversion, power splitting, core-to-core switching and combination, along with remote probe characterization.

Unlike the system we have recently introduced in Ref [25], this platform operates with an arbitrary weak probe. This results in fundamentally different spatiotemporal dynamics, suitable for low-power applications. Once the BCB mode state is set by the launch conditions, the BCB power can be tuned for on-demand reconfiguration of the probe.

Our experimental results are supported by a theoretical model that aligns with the experimental findings and extends to MMFs and MCFs with an arbitrary number of modes and cores.

These results introduce a major shift in critical applications whose tunability currently relies on electro-optical or thermo-optical modulation, offering a faster and more energy-efficient alternative through all-optical manipulation, a keystone for future reconfigurable optical networks and optical computing.

Among these applications, mode conversion is crucial for space-division-multiplexing [32,33]. Our platform enables not only full mode-to-mode conversion in the output probe, but more generally to achieve a tuneable combination of modes (Fig. 3). This latter capability is essential for broadband nonlinear applications[34] and multimode interferometry[35].

Power splitting underpins power delivery, optical feedback and network access [36,37]. The ability to achieve all-optically an arbitrary splitting ratio (Fig. 4a) represents a crucial step towards real-time optimization in time-varying scenarios such as transparent optical networks.

As for our outcomes on core-to-core switching and combining (Fig.4b,c and Fig. 5), these promise advancements in high-speed data transmission. Current switching systems are based on external devices connected to network fibres[38–40], increasing cost and complexity of the design, latency, and overall insertion losses. On the other hand, our approach suggests the feasibility of all-optical tuneable core-to-core switching directly within multicore fibres at sub-nanosecond timescale, paving the way for seamless fibre transmission through compact, all-fibre based ultrafast switchers.

Lastly, probe remote characterization (Fig. 6) offers a novel scenario of applicability, allowing for real-time monitoring of fibre parameters or complex multimode optical signals from remote locations.

The implementation of these operations in a single platform underscores its versatility, a critical feature of next-generation photonic systems[29]. Two further points merit discussion. First, our analysis suggests that the ultimate switching time could be sub-picosecond, therefore beyond the reach of any electronic system. Moreover, scaling these results to highly nonlinear materials, such as silicon or silicon nitride, promises further reductions in power consumption and size. Our results in highly nonlinear fibre (Supplementary Information 2) support this hypothesis.

We have recently demonstrated our ability to design and control coupling in arrays of coupled waveguides[41], which represent the counterpart of MCFs on-chip. In this framework, light-by-light manipulation of the probe would add a critical degree of control for ultrafast reconfiguration. This paves the way for programmable photonics circuits[30,42,43] (hosted on MCFs, on-chip or hybrid) that are all-optically reconfigurable and where basic logic blocks, like the DCF with integrated BCB, are cascaded to implement complex operations.

**METHODS**

**Theoretical framework.** We consider two counter-propagating beams in a polarization-maintaining multimode (or multicore) optical fibre of length $L$ supporting $N$ guided spatial modes. If the beams are co-polarized along the p-axis (p is one of the birefringence axes) and are centred at the same carrier wavelength $\lambda$, their spatio-temporal dynamic is described by the following set of coupled nonlinear Schrödinger equations(CNLSEs) [25]:

$$\partial_z f_n + v_n^{-1} \partial_t f_n = -i\gamma_{nn}|f_n|^2 f_n + if_n \sum_{m=1}^{N} \gamma_{nm}(\kappa|b_m|^2 + 2|f_m|^2) + i\kappa b_n^* \sum_{\substack{m=1 \\ m \neq n}}^{N} \gamma_{nm} b_m f_m$$

$$-\partial_z b_n + v_n^{-1} \partial_t b_n = -i\gamma_{nn}|b_n|^2 b_n + ib_n \sum_{m=1}^{N} \gamma_{nm}(\kappa|f_m|^2 + 2|b_m|^2) + i\kappa f_n^* \sum_{\substack{m=1 \\ m \neq n}}^{N} \gamma_{nm} b_m f_m$$

(4)

Here $\kappa = 2$, while $f_n(z,t)$ and $b_n(z,t)$ indicate the slowly varying amplitudes of the forward and backward mode $n$, respectively. Equation (4) is completed with the boundary conditions that define the input fields, namely $f_n(0,t)$ and $b_n(L,t)$. The instantaneous amplitudes $\tilde{f}_n$ and $\tilde{b}_n$ are related to the slowly varying amplitudes through $\tilde{f}_n = f_n exp(-i\beta_{np}z)$ and $\tilde{b}_n = b_n exp(i\beta_{np}z)$, with $\beta_{np}(\lambda)$ the propagation constant of the p-polarized mode $n$ at wavelength $\lambda$. For the purposes of our subsequent analysis, it is useful to rewrite the relation between $\tilde{f}_n$ and $f_n$ in matrix form, namely $\tilde{\boldsymbol{F}} = \boldsymbol{E_\beta} \boldsymbol{F}$, where $\tilde{\boldsymbol{F}}$ and $\boldsymbol{F}$ are 1x$N$ vectors with elements $\tilde{f}_n$ and $f_n$, respectively, whereas $\boldsymbol{E_\beta}$ is the diagonal matrix with entries $\boldsymbol{E_\beta}[n,n] = exp(-i\beta_{np}z)$. The coefficients $v_n$ and $\gamma_{nm}$ in equation (4) are the group velocity of mode $n$ and the Kerr coefficient for the nonlinear interaction between mode $n$ and $m$ [44], respectively. They are computed via finite-element-method software (see Supplementary Information 1) after measuring the refractive index profile with an optical fibre analyser.

Group velocity dispersion (GVD) and higher-order dispersion terms are ignored in equation (4) as the corresponding characteristic lengths are substantially larger than the fibre lengths used in our experiments.

According to the normalization of the coefficients in equation (4), $|f_n(z,t)|^2$ and $|b_n(z,t)|^2$ indicate the instantaneous power coupled to the forward and backward mode $n$, respectively. The total forward energy, $\int_t \sum_n |f_n|^2\, \partial t$, and backward energy, $\int_t \sum_n |b_n|^2\, \partial t$, are conserved except for propagation losses, which are negligible in the short fibres used.

The last summation on the right-hand-side of equation (4) describes the intermodal power exchange between forward and backward modes. A key feature of our counterpropagating setup is that, because the forward and backward beams are co-polarized and centred at the same carrier wavelength, each component of this summation is automatically phase-matched, irrespectively of the carrier wavelength and the fibre parameters. Consequently, a nonlinear dynamic is triggered where all modes can simultaneously exchange energy, rather than just a single pair of phase-matched modes, as typically occurs in co-propagating setups.

If forward and backward beams are orthogonally polarized along different birefringence axes, the nonlinear intermodal interaction is reduced by a factor of 1/3 ($\kappa=2/3$ in equation (4)), and each component of the last summation is subject to a polarization phase-mismatch $\Delta\beta = \beta_{nx}(\lambda) - \beta_{ny}(\lambda) + \beta_{my}(\lambda) - \beta_{mx}(\lambda)$. However, in the fibres under test, this phase mismatch barely impacts the dynamic, since the corresponding beat length $2\pi/\Delta\beta$ is typically larger than the interaction length $L_{in}$ between forward and backward beams. The latter equals the fibre length in the continuous-wave (CW) case, while reads $L_{in} = \tau_p c$ in pulsed operation, where $\tau_p$ is the pulse width of the forward and backward beams and $c$ is the velocity of light in the fibre. Similarly, if forward and backward beams are centred at different carrier wavelengths $\lambda_f$ and $\lambda_b$, the induced phase-mismatch is negligible whenever the detuning $\Delta\lambda = |\lambda_f - \lambda_b| \ll \lambda_0^2/(c\, L_{in} |v_n^{-1} - v_m^{-1}|)$ with $\lambda_0 = (\lambda_f + \lambda_b)/2$ [25]. This enables tuning of the wavelength selectivity for applications such as core-to-core switching, which occurs only when the probe wavelength is sufficiently close to the BCB wavelength ($\Delta\lambda < 10$ nm in the fibres under test).

**System linearization: probe-BCB equations.** In the following, in accordance with the notation used in the manuscript, we indicate the forward and backward beam with probe and BCB, respectively. Let us consider the continuous-wave case in which the probe is a signal with low power. Equation (4) is reduced to equation (5) by using a perturbation approach where the less significant nonlinear terms are ignored along with time-varying terms ($\partial_t f_n$ and $\partial_t b_n$):

$$\partial_z f_n = +i f_n \sum_{m=1}^{N} \gamma_{nm}\kappa |b_m|^2 + i\kappa b_n^* \sum_{\substack{m=1\\m\neq n}}^{N} \gamma_{nm} b_m f_m$$

$$-\partial_z b_n = i\theta_n b_n$$

(5)

Here $\theta_n = -\gamma_{nn}|b_n|^2 + \sum_{m=1}^{N} 2\gamma_{nm}|b_m|^2$ plays the role of a nonlinear phase shift induced by self-phase and cross-phase modulation. The solution for the BCB mode $n$ reads as $b_n(z) = b_n(0)exp(-i\theta_n z)$, therefore its amplitude is preserved in propagation, except for the nonlinear phase variation. We insert this solution in the first line of equation (5) and we use the transformation $f_n = \widehat{f_n} exp(i\theta_n z)$. This latter transformation can be recast in matrix form as $\boldsymbol{F} = \boldsymbol{E_\theta}\widehat{\boldsymbol{F}}$, where $\widehat{\boldsymbol{F}}$ is the 1x$N$ vector with elements $\widehat{f_n}$ and $\boldsymbol{E_\theta}$ is the diagonal matrix whose entry $\boldsymbol{E_\theta}[n,n] = exp(i\theta_n z)$. We finally obtain a system of linear differential equations (LDE) for $\widehat{f_n}$ that can be written as $\partial z\, \widehat{\boldsymbol{F}} = i\, \boldsymbol{A}\, \widehat{\boldsymbol{F}}$, where $\boldsymbol{A}$ is the $N$x$N$ matrix whose diagonal elements $\boldsymbol{A}[n,n] = -\theta_n + \kappa \sum_{m=1}^{N} \gamma_{nm}|b_m|^2$, and $\boldsymbol{A}[n,m] = \kappa\gamma_{nm}b_m(0)b_n(0)^*$ for $n\neq m$. The matrix $\boldsymbol{A}$ stores therefore the information on the BCB mode state. The solution to the above-mentioned LDE system is readily found by eigenvector decomposition of matrix $\boldsymbol{A}$, namely $\widehat{\boldsymbol{F}}(L) = \boldsymbol{V}exp(i\boldsymbol{\Lambda} L)\boldsymbol{V}^{-1}\widehat{\boldsymbol{F}}(0)$, where $\boldsymbol{V}$ and $\boldsymbol{\Lambda}$ are the matrix of eigenvectors and eigenvalues of $\boldsymbol{A}$, respectively. Now, by making use of the relations previously introduced, namely $\boldsymbol{\mathcal{F}} = \boldsymbol{E_\beta}\,\boldsymbol{F}$ and $\boldsymbol{F} = \boldsymbol{E_\theta}\widehat{\boldsymbol{F}}$, we derive the solution $\boldsymbol{\mathcal{F}}(L) = \boldsymbol{M}\boldsymbol{\mathcal{F}}(0)$ previously indicated as equation (1), where $\boldsymbol{M} = \boldsymbol{E_\beta}\,\boldsymbol{E_\theta}\boldsymbol{V}exp(i\boldsymbol{\Lambda} L)\boldsymbol{V}^{-1}$, while $\boldsymbol{\mathcal{F}}(0) \equiv \boldsymbol{\mathcal{F}_{in}}$ and $\boldsymbol{\mathcal{F}}(L) \equiv \boldsymbol{\mathcal{F}_{out}}$ are the input and output probe mode state, respectively.

The above-mentioned solution is generally applicable to any multimode fibre system, including coupled multicore fibres. In the latter case, it is useful to derive a relationship between the field in the individual cores of the fibre. We proceed by using a couple mode theory approach, where the modes of the multicore fibre are approximated as a linear combination of the fields in the cores, namely, $\boldsymbol{\mathcal{F}_c} = \boldsymbol{T}\boldsymbol{\mathcal{F}}$, where $\boldsymbol{T}$ is a transformation matrix and $\boldsymbol{\mathcal{F}_c}$ is the 1x$N$ vector whose element $f_{c,n}$ indicates the field in the core $n$. In the simplest case of a DCF with single-mode cores, the two guided modes are well approximated as the in-phase and anti-phase sum of the fields in the cores, therefore $\boldsymbol{T} = [1,1;1,-1]/\sqrt{2}$. In general, the unitary $\boldsymbol{T}$ matrix strictly depends on the core-to-core arrangement. In the case of the TCF under test (corresponding results are illustrated in Fig. 5b-d), where the cores are arranged at the vertices of an isosceles triangle with 30-deg base angle and ~16.5 µm base, we have $\boldsymbol{T} = [\sqrt{2}, 0, \sqrt{2}; 1, \sqrt{2}, -1; 1, -\sqrt{2}, -1]/2$.

When the probe is in linear regime, the solution of the full CNLSEs equation (4) yields the same results as the simplified system equation (5) and the analytical formulas equations (1) and (2), confirming the validity of our model. The advantage of using equations (1) and (2) is that they directly provide the probe mode/core state as a function of matrices $\boldsymbol{M}$ and $\boldsymbol{T}$, eliminating the need for propagation codes. Notably, equations (1) and (2) allow identifying the optimal matrix $\boldsymbol{M}$, and then the related optimal BCB mode state, to implement the all-optical applications introduced in this work.

**Relation between probe and BCB mode/core states and mode/core power distribution.** The matrix $A$ can be decomposed as $A = E^*_{\angle B_0} A' E_{\angle B_0}$, where $E_{\angle B_0}$ is the diagonal matrix whose entry $E_{\angle B_0}[n, n] = exp(-i \arg(b_n(0)))$ identifies the phase of the BCB mode $n$ in z=0, and $A'$ is the matrix created from $A$ by replacing the non-diagonal entries $\kappa \gamma_{nm} b_m(0) b_n(0)^*$ with the corresponding magnitude $\kappa \gamma_{nm} |b_m(0)||b_n(0)|$. Matrices $A$ and $A'$ are therefore equivalent except for the phase information of the BCB, which is missing in $A'$.

By exploiting the above-mentioned decomposition, the relation $\partial z \widehat{F} = i A \widehat{F}$ can be rewritten as $\partial z (E_{\angle B_0} \widehat{F}) = i A' (E_{\angle B_0} \widehat{F})$, meaning that the dynamics of the transformed vector $E_{\angle B_0} \widehat{F}$ depends solely on the modified matrix $A'$. Since $|E_{\angle B_0} \widehat{F}| = |\widehat{F}| = |\mathcal{F}|$, we conclude that the probe mode power distribution $|\mathcal{F}|$ is fully determined by $A'$, rather than $A$. In other words, the output probe mode power distribution only depends on the BCB mode power distribution (that is preserved in propagation and is fixed by the launch conditions), but not on the BCB mode relative phases. This is not generally true for the output probe core power distribution, which depends instead on the full BCB mode state.

**Experiments.** In our experiments, the BCB operates in a pulsed configuration rather than as a CW, which enhances the peak power and, consequently, the system nonlinearity. The probe could in principle operate in the CW regime with indefinitely low power. In practice, the probe-to-BCB power imbalance in our experiments is ~1:20. Indeed, a lower probe power would result in a weak signal-to-noise ratio, thus degrading the image quality, and preventing an accurate mode decomposition of the probe. Consequently, both BCB and probe are pulsed in our experiments (which does not change the main outcomes, see Supplementary Information 5). Specifically, 0.5 ns-pulsed probe and BCB are generated by splitting the beam from an in-house built linearly polarized ytterbium master oscillator power amplifier having central wavelength λc=1040nm and a repetition rate of 800 kHz [45]. The probe and BCB are then injected at the two opposite ends of the fibres under test. Four distinct fibres are employed (see Supplementary Information 1): a polarization-maintaining (PM) few-mode fibre (PM1550-xp from Thorlabs) supporting 3 guided modes at λc; a highly nonlinear PM few-mode fibre (PMHN1 from Thorlabs) supporting 3 guided modes at λc; and then a homemade dual core fibre (DCF) and three-core fibre (TCF) supporting respectively 2 and 3 guided modes at λc.

The input power and polarization of probe and BCB are controlled with a proper combination of polarization beam splitters and half-wave-plates (HWP$_2$ to HWP$_5$ in Fig. 2). By adjusting the phase pattern displayed on the screen of a spatial light modulator, we control the mode state of the input BCB, namely, its power distribution and relative phase over the fibre modes. A spatial phase plate is used to excite an arbitrary combination of modes at the probe input end for the PM1550-xp and PMHN1, while the input probe is selectively coupled into a single core to excite a combination of modes in the DCF and TCF.

The test fibre at the BCB input end is cleaved with an angle of 8-deg to eliminate back reflection of the BCB, whereas the probe input end is perpendicularly cleaved to ensure high-quality mode excitation. The output probe is sampled using a wedge with an incident beam angle of ~10-deg, ensuring that the sampled beam preserves the output probe polarization. The near-field and far-field intensity profiles of probe and BCB are measured with infrared cameras, with the output probe profiles corrected by subtracting the BCB reflection from the flat-cleaved fibre end. Mode decomposition of the probe and BCB is then implemented based on the measured intensity profiles. Specifically, a reconstructed spatial distribution is generated by numerically determining the mode state through an iterative process, where the Stochastic Parallel Gradient Descent algorithm[46] is successfully applied. The reconstructed distribution typically exhibits a correlation as high as 99%[47–49] with the measured spatial profile, which confirms the effectiveness of the mode decomposition method.

In the MCFs under test, the power in each individual core is measured by integrating the intensities within the core areas in the near-field intensity profiles. To analyse the temporal evolution of core-to-core power switching, the output probe pulses from each core are characterized by an oscilloscope. As shown in Fig. 2, the output probe is imaged at the pinhole position via a pair of lenses (focal lengths=13.86 mm and 500 mm, providing a magnification factor of ~36x). With a clear aperture of ~200 μm, the pinhole can effectively filter out the beam from a single core. The filtered output probe is then coupled through a telescope into a multimode fibre connected to the oscilloscope, with a replica imaged onto the camera using another telescope. Due to the flat cleave at the input-probe fibre end, the BCB reflection at this facet propagates in the same direction of the probe and can then also be measured (see BCB reflection in Fig. 4e). However, the BCB reflected pulses are separated from the output probe pulses due to the differential travelling path length, with a delay essentially determined by the fibre length (~2 ns in Fig. 4e).

# SUPPLEMENTAL DOCUMENT:
# Sub-nanosecond all-optically reconfigurable photonics in optical fibres

**KUNHAO JI[1,*], DAVID. J. RICHARDSON[1,2], STEFAN WABNITZ[3], AND MASSIMILIANO GUASONI[1,*]**

[1]*Optoelectronics Research Centre, University of Southampton, Southampton SO17 1BJ, United Kingdom*
[2]*Microsoft (Lumenisity Limited), Unit 7, The Quadrangle, Abbey Park Industrial Estate, Romsey, SO51 9DL, United Kingdom*
[3]*Department of Information Engineering, Electronics and Telecommunications (DIET), Sapienza University of Rome, 00184 Rome, Italy*
*\*m.guasoni@soton.ac.uk*

## Supplementary Information 1: Fibre modes and coefficients

In Fig. S1, we present the spatial profiles of the modes of the fibres tested in our experiments: a homemade dual-core fibre (DCF) and three-core fibre (TCF) supporting respectively 2 and 3 guided modes; and then a polarization-maintaining (PM) few-mode fibre (PM1550-xp from Thorlabs) and a highly nonlinear PM few-mode fibre (PMHN1 from Thorlabs) supporting 3 guided modes. Note that, in our experiments, the DCF is used both as bimodal fibre to illustrate multimode manipulation (Figs. 3 and 6 of the manuscript) and as multicore fibre for multicore manipulation (Fig. 4 of the manuscript). Modes $M_1$ and $M_2$ in the DCF correspond to the supermodes with the core fields in phase and anti-phase, respectively. Modes $M_1$, $M_2$ and $M_3$ in the PM1550-xp and PMHN1 fibre correspond to the standard linearly polarized modes $LP_{01}$, $LP_{11e}$, and $LP_{11o}$, respectively. Note that in these fibres $LP_{11e}$ and $LP_{11o}$ are non-degenerate.

Tables S1 and S2 list the Kerr coefficients and inverse group velocities for the fibres under test. These coefficients, along with the spatial profiles of the modes, were computed using finite element method simulations (central wavelength $\lambda_c$ =1040 nm).

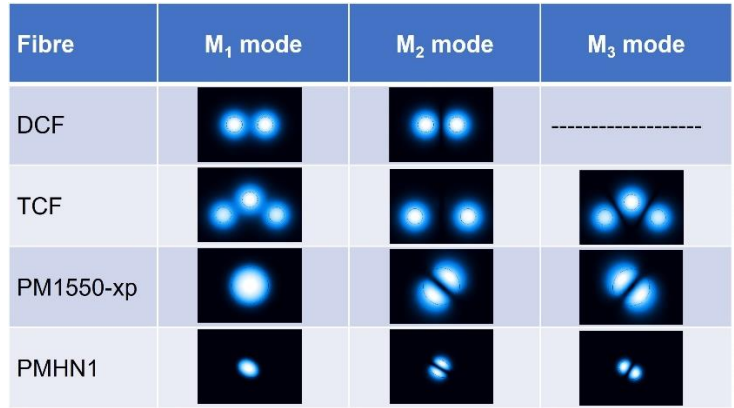

**Fig. S1. Modes of the fibres under test.** Spatial distribution (intensity) computed from finite element method simulations.

**Table S1: Kerr coefficients**

| Fibre | $\gamma_{11}$(W$^{-1}$km$^{-1}$) | $\gamma_{22}$(W$^{-1}$km$^{-1}$) | $\gamma_{33}$(W$^{-1}$km$^{-1}$) | $\gamma_{12}=\gamma_{21}$(W$^{-1}$km$^{-1}$) | $\gamma_{13}=\gamma_{31}$(W$^{-1}$km$^{-1}$) | $\gamma_{23}=\gamma_{32}$(W$^{-1}$km$^{-1}$) |
|---|---|---|---|---|---|---|
| DCF | 3.00 | 3.12 | --- | 3.06 | --- | --- |
| TCF | 2.30 | 3.17 | 2.46 | 1.56 | 2.37 | 1.61 |
| PM1550-xp | 3.00 | 2.59 | 2.44 | 1.73 | 1.67 | 0.84 |
| PMHN1 | 13.20 | 12.50 | 10.71 | 8.41 | 7.32 | 3.84 |

**Table S2: Inverse group velocity $v_n^{-1}$ of mode-n**

| Fibre | $v_1^{-1}$ (ns/m) | $v_2^{-1} - v_1^{-1}$ (ps/m) | $v_3^{-1} - v_1^{-1}$ (ps/m) |
|---|---|---|---|
| DCF | 4.906 | 1.068 | ------ |
| TCF | 4.906 | 0.714 | 1.513 |
| PM1550-xp | 4.895 | 0.452 | 0.516 |
| PMHN1 | 4.978 | 11.919 | 0.744 |

## Supplementary Information 2: Tuneable mode manipulation in three-mode fibres

Mode manipulation has been tested experimentally in two three-mode commercial fibres from Thorlabs, namely PM1550-xp and PMHN1. Both these fibres support the propagation of $LP_{01}$, $LP_{11e}$ and $LP_{11o}$ modes at a wavelength $\lambda_c$=1040 nm. The length of the test fibres is ~0.4m, and the probe and BCB are co-polarized.

Fig. S2a-c display the output probe mode distribution as a function of the BCB peak power in the PM1550-xp fibre and in three distinct cases. By adjusting the BCB mode distribution (reported at the top of each panel) we trigger different dynamics in the output probe. In Fig.S2a, the power exchange between the $LP_{01}$ and $LP_{11o}$ modes is promoted. For a BCB peak power of ~11kW, the output probe exhibits approximately equal power distribution between the $LP_{01}$ and $LP_{11o}$ modes. On the contrary, in Fig.S2b, the power exchange between the $LP_{01}$ and $LP_{11e}$ modes is favoured. Finally, in Fig.S2c we observe power exchange between all the 3 modes. Now, for a BCB peak power of ~11 kW, the output probe exhibits approximately equal power distribution between all the 3 modes.

Fig. S2d,e display the results in the PMHN1 fibre in two distinct instances. The input probe mode state is similar in both cases, and BCB mode distribution is properly adjusted to trigger different dynamics, with most of the output probe power redirected either on mode $LP_{01}$ (Fig. S2d) or $LP_{11o}$ (Fig. S2e). It is worth noting that the BCB peak power required to achieve relevant dynamics is substantially lower than in the case of the PM1550-xp, being as small as ~1kW. This is essentially due to the high nonlinearity of the fibre. The Kerr coefficients are indeed substantially larger in the PMHN1 fibre than in the PM1550-xp fibre (see Table S1).

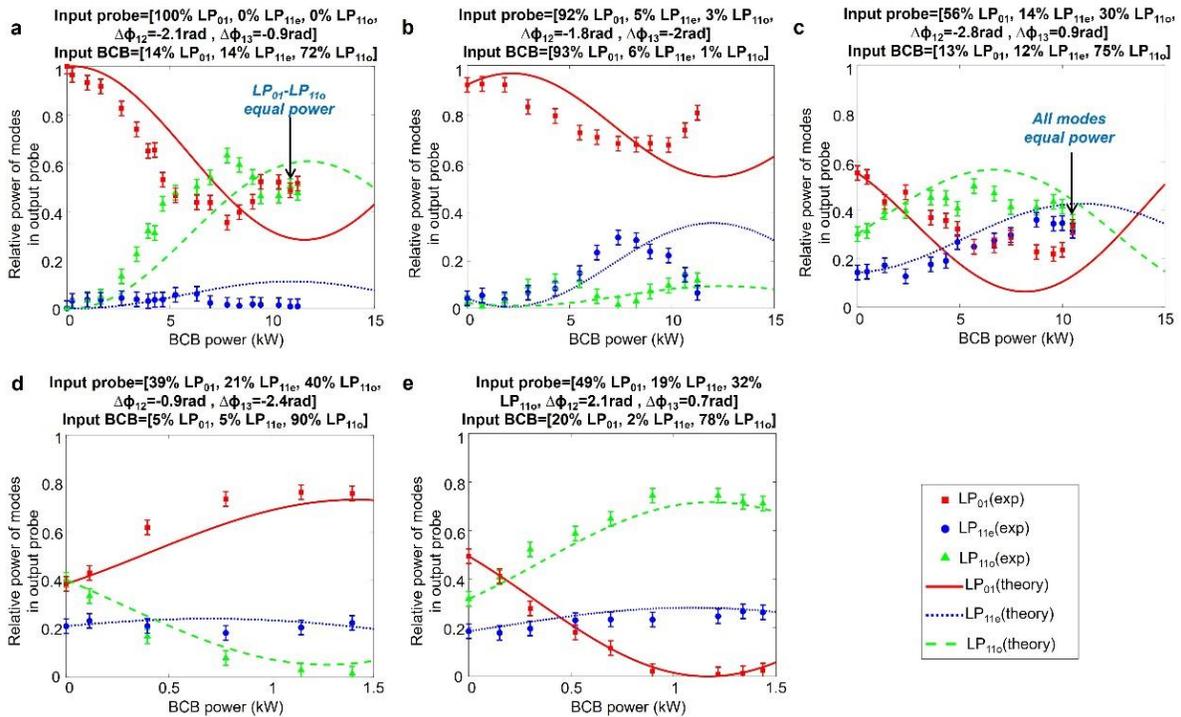

**Fig. S2. Tuneable mode manipulation in three-mode fibres (PM1550-xp and PMHN1). a-c.** Mode decomposition of the output probe as a function of the BCB peak power for the PM1550-xp fibre. Input probe mode state and BCB mode distribution are reported at the top of each panel. **d,e.** Same as panel a-c but for the HNF1 fibre. Exp=experimental results; Theory= theoretical results from solution of equation (1) in the main manuscript. Error bars of ±3% are added to the measured relative power of each mode, which represents the estimated uncertainty of our mode decomposition algorithm.

## Supplementary Information 3: Mode manipulation via probe-BCB relative polarization

The relative polarization between input probe and BCB represents a further parameter to all-optically reconfigure the output probe. Fig. S3 illustrates some experimental results in the case of a bimodal fibre (DCF). The evolution of the output probe mode distribution as a function of the BCB power is reported when input probe and BCB are either co-polarized (Fig. S3a) or orthogonally polarized (Fig. S3b) and while maintaining the same input probe and BCB mode state. In the case of orthogonal polarization, we observe a slower modal conversion dynamic, due to the weaker interaction between the probe and the BCB (coefficient $\kappa$ is reduced by a factor of 1/3 in equation (4) of the manuscript). Fig. S3c shows the mode distribution of the output probe when the probe-BCB relative polarization is continually adjusted from 0 deg (co-polarized) to 90 deg (orthogonally polarized), whereas the BCB power is fixed (8 kW peak-power). We observe that the fraction of the output probe coupled to mode $M_1$ ($M_2$) is tuneable in the range 37-75% (30-70%).

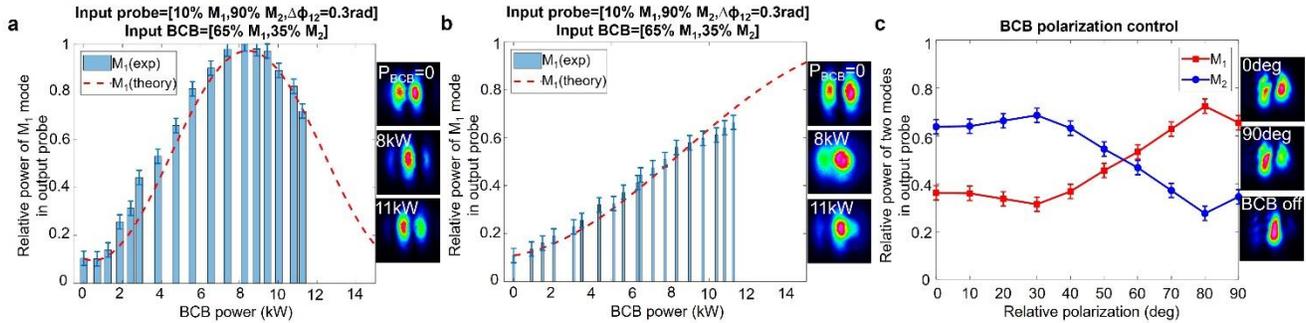

**Fig. S3. Tuneable mode manipulation by adjusting the polarization. a,b**. Comparison between output probe mode distribution in the co-polarized case (panel a) and orthogonally-polarized case (panel b). The insets show the far-field intensity of the output probe at different BCB power $P_{BCB}$. **c**. Mode distribution of the output probe as a function of the probe-BCB relative polarization for a fixed BCB power (8kW peak power). The insets show the far-field intensity at 0 deg, 90 deg and when BCB is turned off.

## Supplementary Information 4: Linear VS nonlinear probe regime

The dynamics of the counter-propagating system, as described in equation (4) of the manuscript, are significantly influenced by the degree of nonlinearity of both the probe and BCB.

When both beams operate in a strongly nonlinear regime, the system exhibits asymptotic attraction to or rejection of specific mode states, as reported in Ref. S1. For instance, in the case of a bimodal fibre, the output probe beam is attracted towards the mode state orthogonal to the input BCB, and vice versa. In the example shown in Fig. S4a,b, we simulate a bimodal fibre with parameters $L$=1m, $\gamma_{11} = \gamma_{12} = \gamma_{22} = 1/W/km$. The input probe beam is entirely coupled to mode $M_1$, while the input BCB is distributed with 60% of its power in mode $M_1$ and 40% in mode $M_2$. The probe beam, with a total fixed peak power $P_p = 10$ kW, operates in a highly nonlinear regime (number of nonlinear lengths $L\gamma P_p = 10$, $\gamma = 1/W/km$ being the average Kerr coefficient). As the BCB's peak power increases from 0 to 10 kW, entering itself a strongly nonlinear regime, the mode attraction process outlined above occurs. Indeed, the output probe (Fig. S4a) tends to approach the mode state orthogonal to the input BCB, namely, ~ 40% on mode $M_1$ and ~ 60% on mode $M_2$. In turn, the output BCB (Fig. S4b) tends to approach the mode state orthogonal to the input probe, namely, all power coupled to mode $M_2$.

However, when the probe operates at a low peak power level, therefore remaining in a linear regime (which is the condition underlying the outcomes reported in the manuscript) the dynamics change drastically. The mode attraction process is not triggered. This is shown in Fig. S4c,d where the probe peak power is now arbitrary low (here $P_p$ =0.01 kW, therefore the number of nonlinear lengths $L\gamma P_p = 0.01$). In this case, the output BCB's mode composition remains unchanged, mirroring the input (Fig. S4d). Meanwhile, the output probe mode distribution exhibits a sinusoidal evolution as the BCB power increases (Fig. S4c), in line with the predictions of our theoretical model (equation (1) in the manuscript) and the experimental outcomes reported in the manuscript.

The results shown in Fig. S4 are generalizable to fibres with arbitrary coefficients. As a rule of thumb, if the number of nonlinear lengths exceeds 5, then the probe operates in a strongly nonlinear regime, as depicted in Fig. S4a,b. Conversely, if it is below 0.5, then the probe operates in a linear regime, as illustrated in Fig. S4c,d.

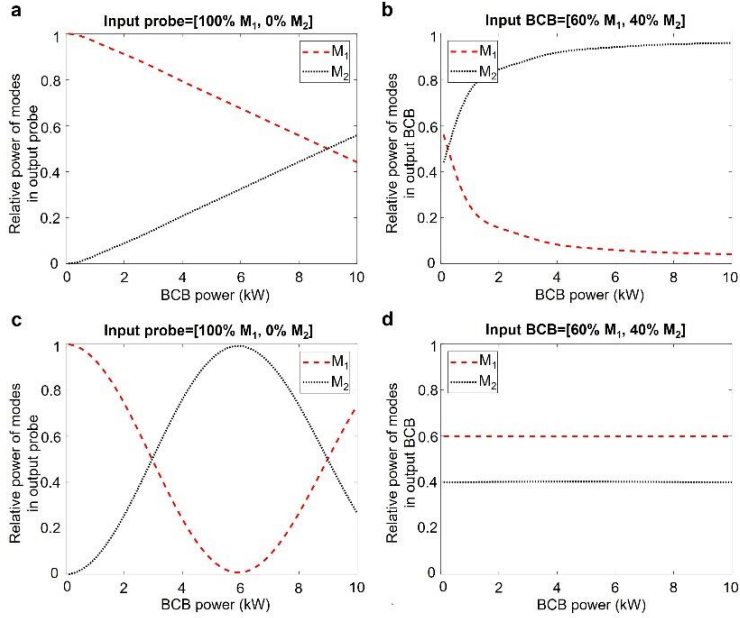

**Fig. S4. Comparison between high power probe (mode attraction) and low power probe in a bimodal fibre. a-b**. Mode distribution of the output probe (a) and output BCB (b) versus the BCB peak power when the probe is in a strong nonlinear regime (peak power fixed to 10 kW). The output probe is asymptotically attracted to the mode state orthogonal to the input BCB, and viceversa. **c-d**. Mode distribution of the output probe (c) and output BCB (d) versus the BCB peak power when the probe is in linear regime (peak power fixed to 0.01 kW). The output probe mode distribution oscillates sinusoidally as a function of the BCB power, whereas the BCB mode distribution is unchanged.

## Supplementary Information 5: Ultrafast dynamics

In our experiments, we have demonstrated that the core-to-core power ratio of the output probe can be switched on a sub-nanosecond timescale (see Fig. 4c,d of the manuscript).

In the following, we discuss some numerical results that illustrate the core-to-core switching mechanism in detail. These results are obtained by simulation of the full CNLSEs reported in equation (4) of the manuscript. We focus on the case of pulsed BCB. For simplicity, we illustrate the case of a dual core fibre with $L$=1 m, $\gamma_{12} = 1/W/km, \gamma_{11} = \gamma_{22} = 2/W/km$. The system under analysis possesses three critical timescales: the BCB pulse width $\tau_P$; the BCB repetition rate R; and the time of flight $\tau_F = L/c$ in the fibre (c= light velocity in the fibre). In these simulations, the BCB is equally distributed in the 2 fibre modes and the BCB pulse width $\tau_p$=1 ps. Full core-to-core conversion (from 100/0 to 0/100) in the output probe is achieved for a BCB peak power $P_{BCB}$ =6.2 MW. The probe peak power is instead arbitrary low.

Initially, we consider the case of a continuous-wave probe. In Fig. S5, we show the output probe temporal dynamics when the BCB is respectively off (Fig. S5b) and on (Fig. S5c). When the BCB is off, the output probe power is fully coupled to core 1 (100/0 power ratio). When the BCB is on, the first BCB pulse with $P_{BCB}$ =6.2 MW (BCB pulse 1) triggers the full switching of the output probe power ratio from 100/0 to 0/100. This state is maintained over a time window $2\tau_F = 2L/c$, after which the power ratio returns to 100/0 (BCB off condition).

At the following BCB pulse (BCB pulse 2), after a time 1/R, the dynamics repeat. However, the peak power of BCB pulse 2 is now lower, i.e. 3.2 MW, resulting in a reduced conversion (45/55 power ratio). A similar dynamic applies to BCB pulse 3, whose low power results in a weak conversion (80/20 power ratio). Note that if the condition 1/R =$2\tau_F$ is met, then it is possible to maintain the output probe power ratio 0/100, as shown in Fig. S6.

Similar considerations apply when the probe is pulsed: regardless of the probe pulse width, each probe pulse within a time window of width $2\tau_F$ is switched, as reported in Fig. S7.

It is worth noting that the results illustrated above are generalizable to fibres with different parameters and/or more cores, as well as different pulse widths. Similarly, these results would extend to other waveguide systems beyond optical fibres. For example, using silicon-based integrated waveguides (which possess nonlinearities > 3 orders of magnitude higher than standard optical fibres), similar results to Figs. S5-S7 could be achieved but reducing the peak powers in the fraction-of-kW range. These peak power levels, along with ps pulses and GHz repetition rates, can be currently obtained using commercial fibre lasers. The combination of innovative all-optical pulse generation techniques (see Ref. S2) with our all-optical switching approach would underpin the development of an ultrafast all-optical integrated switching system.

Note finally that the switching time $\tau_{switch}$, required to complete the core-to-core switching, is of the order of the BCB pulse width $\tau_p$ (see insets of Fig. S5). The latter is ultimately constrained by distortion due to interplay among dispersion and self-phase modulation, which poses a limit to the minimum BCB pulse width.

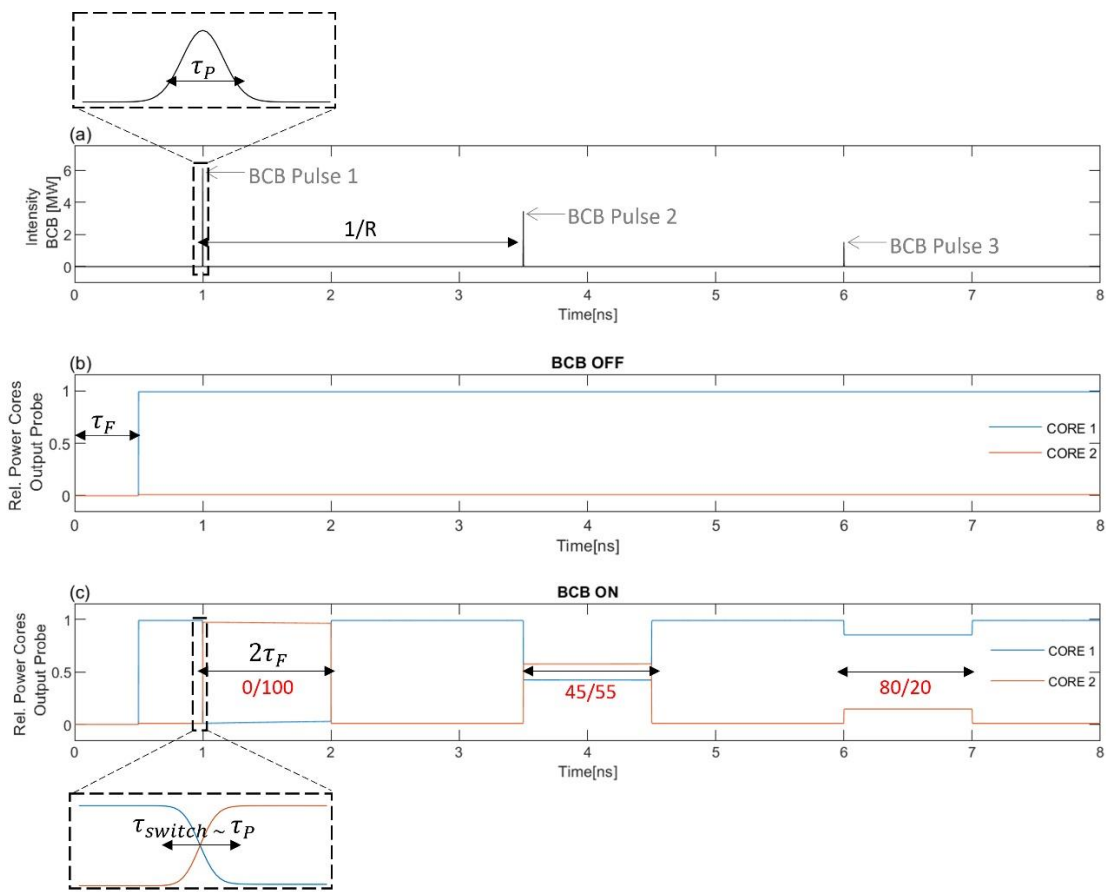

**Fig. S5. Ultrafast dynamics with pulsed BCB, CW probe.** Each BCB pulse (see panel a and related inset) causes a switching in the output probe core-to-core power ratio. The achieved power ratio (here 0/100, 45/55 and 80/20 for BCB pulses 1, 2 and 3, respectively) depends on the BCB pulse peak power. The switching is preserved over a time window of length $2\tau_F$, where the time of flight $\tau_F$ is shown in panel b.

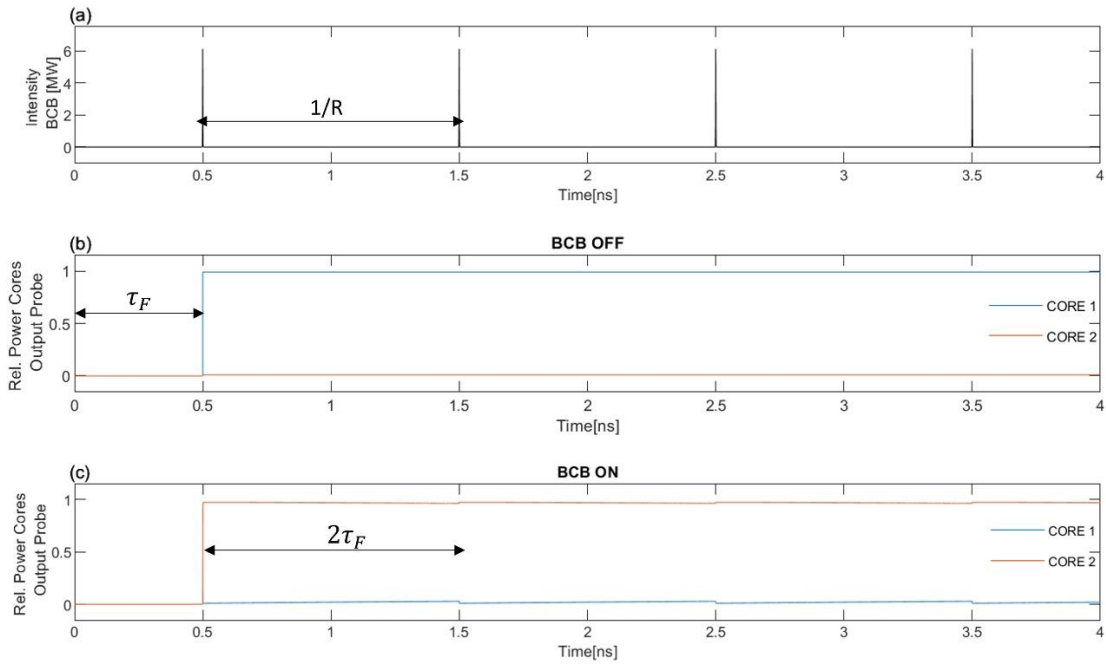

**Fig. S6. Preserving 0/100 state.** As Fig. S5, but here each BCB pulse has peak power 6.2 MW and the condition $1/R = 2\tau_F$ is met. Consequently, the output probe core-to-core power ratio 0/100 is preserved.

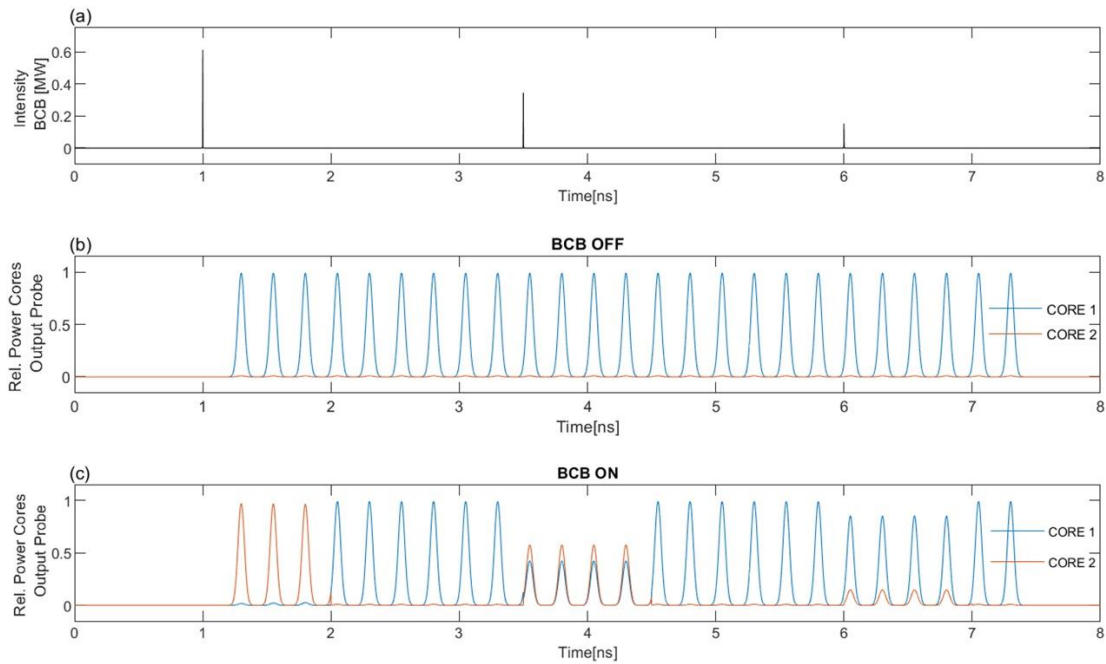

**Fig. S7. Ultrafast dynamics with pulsed BCB, pulsed probe.** As Fig. S5, but here the probe is pulsed with pulse width =200ps.